\documentclass[aps,prb,twocolumn,groupedaddress]{revtex4}

\usepackage{graphicx}

\begin{document}

\title{Bulk 5$f$ electronic states in uranium monochalcogenide US}

\author{Y. Takeda}
\email[]{ytakeda@spring8.or.jp}
\affiliation{Japan Atomic Energy Agency, Synchrotron Radiation Research Center SPring-8, Sayo, Hyogo 679-5148, Japan}
\author{T. Okane}
\affiliation{Japan Atomic Energy Agency, Synchrotron Radiation Research Center SPring-8, Sayo, Hyogo 679-5148, Japan}
\author{T. Ohkochi}
\affiliation{Japan Atomic Energy Agency, Synchrotron Radiation Research Center SPring-8, Sayo, Hyogo 679-5148, Japan}
\author{Y. Saitoh}
\affiliation{Japan Atomic Energy Agency, Synchrotron Radiation Research Center SPring-8, Sayo, Hyogo 679-5148, Japan}
\author{H. Yamagami}
\affiliation{Japan Atomic Energy Agency, Synchrotron Radiation Research Center SPring-8, Sayo, Hyogo 679-5148, Japan}
\affiliation{Department of Physics, Kyoto Sangyo University, Kyoto 603-8555, Japan}
\author{A. Fujimori}
\affiliation{Department of Physics, The University of Tokyo, Hongo, Tokyo 113-0033, Japan}
\author{A. Ochiai}
\affiliation{Center for Low Temperature Science, Tohoku University, Sendai, Miyagi 980-8578, Japan}

\date{\today}

\begin{abstract}

We have investigated the electronic states of the uranium monochalcogenide US, which is a typical ferromagnetic uranium compound, using soft x-ray photoemission spectroscopy (SX-PES). 
In early ultraviolet photoemission spectroscopy studies, two peak structures of the U 5$f$ states were observed and have been interpreted that one has an itinerant character around the Fermi level ($E_\mathrm{F}$) and the other located below $E_\mathrm{F}$ has a character of localized U 5$f$ electrons.
In this study, the intrinsic bulk valence-band spectrum of US was first deduced by estimating the contribution of surface states to the valence-band spectrum using core-level photoemission spectra.
We conclude that the electronic structure of US can be basically described by the itinerant nature of the U 5$f$ electrons from comparison with theoretical valence-band spectra obtained by band-structure calculation in the local-density approximation. 

\end{abstract}


\maketitle

Uranium compounds display unique and attractive properties, for example, showing the coexistence of superconductivity and magnetism derived from U 5$f$ electrons.
It has not been settled how the U 5$f$ electrons play a role in the complex physical properties.
Among the uranium compounds, uranium monochalcogenides (UX$_\mathrm{C}$; X$_\mathrm{C}$ = S, Se, Te) crystallizing in the NaCl structure have been well known as ferromagnetic compounds.
The Curie temperatures ($T_\mathrm{C}$) for US, USe and UTe are 177, 160 and 104 K, respectively \cite{Fournier_1985}.
The values of the $T_\mathrm{C}$ of the UX$_\mathrm{C}$ are high relative to other ferromagnetic uranium compounds \cite{Fournier_1985}.
The lattice constants for US, USe and UTe are 5.489, 5.744 and 6.155 {\AA}, increasing as the X$_\mathrm{C}$ element becomes heavier from S to Te.
As for US, the interaction between atoms is considered to be the strongest.
Since the ferromagnetic state of US is the most stable among the UX$_\mathrm{C}$ compounds, the correct grasp of the electronic structure of US becomes a good starting point to understand the physical properties of the UX$_\mathrm{C}$ as well as other uranium compounds.

The photoemission spectroscopy (PES) technique is a very powerful tool to examine the electronic states in strongly correlated electron systems.
In the 1980's, the electronic states of UX$_\mathrm{C}$ were investigated by vacuum ultraviolet PES (VUV-PES) \cite{Reihl_Common128_1987}. 
Two common U 5$f$ structures were observed in the valence-band spectra of UX$_\mathrm{C}$ taken at $h{\nu}$ = 40 eV. 
One was located around the Fermi level ($E_\mathrm{F}$), referred to as {\it F} structure, and the other was present in the binding energy ($E_\mathrm{B}$) range from 1 to 0.5 eV, referred to as {\it B} structure (see Fig. 2).
It has been interpreted that the {\it F} structure forms the Fermi surface, while the {\it B} structure can be explained by the multiplet structures of U 5$f$${^{2}}$.
Therefore, the coexistence of the {\it B} and {\it F} structures has been interpreted as a dual character that the U 5$f$ electrons show both the itinerant and localized nature.
The peak intensity ratio of the {\it B} structure to the {\it F} structure {\it R}$_{\it B/F}$ increases systematically as the X$_\mathrm{C}$ element goes from S to Te.
The behavior of the two components have been considered as evidence that in going from US to UTe the character of the U 5$f$ electrons changes from the itinerant toward localized one. 
High energy-resolution VUV-PES spectra also showed the clear {\it B} structure at $h{\nu}$ = 40.8 eV using a He discharge lamp \cite{Ito_DoctorThesis} and supported the above interpretation. 
In the recent VUV-PES study for UTe \cite{Durakiewicz_PRL_2004}, the $E_\mathrm{B}$ of the {\it F} structure was shifted toward deeper $E_\mathrm{B}$ on cooling and the {\it B} structure was independent of temperature.
It was pointed out that the change of $E_\mathrm{B}$ of the {\it F} structure can be explained within the mean-field (Stoner-like) theory and concluded that UTe is an itinerant-electron ferromagnet.
This implies that the ferromagnetic properties of UX$_\mathrm{C}$ compounds are derived from the itinerant 5$f$ electrons.
Although the intensity of the {\it B} structure was much larger than that of the {\it F} structure, however, there is no detailed explanation about the origin of the {\it B} structure.
On the other hand, VUV-PES measurements for various actinide compounds have shown that the $E_\mathrm{B}$ of the {\it B} structure depends on materials, and there may be a correlation between the $E_\mathrm{B}$ of the {\it B} structure and the values of the ferromagnetic moments of the materials \cite{Durakiewicz_PRB_2004}.
This relationship implies that the magnetic properties originate from the localized nature of the U 5$f$ electrons.
The $E_\mathrm{B}$ of the {\it B} structure and the magnitude of the ferromagnetic moment of UTe obeys the above correlation too.
Therefore, it has not been yet solved what kind of the character of the U 5$f$ electrons the ferromagnetic properties of UX$_\mathrm{C}$ are originated by.
In addition, it should be noted that the {\it B} structure was observed in VUV-PES spectra of some other actinide compounds \cite{Reihl_PRB_1981, Reihl_PRL_1981, Reihl_PRB_1983, Kumigashira_PRB_2000}.
Most of the PES studies for uranium compounds have been performed by VUV-PES so far.
Generally speaking, since the probing depth of VUV-PES is small, there is a possibility that the VUV-PES spectrum is strongly affected by surface states.
If the surface states are different from the intrinsic bulk states, we have to treat the VUV-PES spectrum very carefully to extract physical interpretation from the spectrum.

In the 4$f$ electron systems, soft x-ray PES (SX-PES) has been very useful to understand the intrinsic bulk electronic states \cite{Laubschat_PRL_1990, Sekiyama_nature}.
In Ce-based compounds, the Ce 4$f{^1}$ and 4$f{^0}$ final states have been observed around $E_\mathrm{F}$ and 1 - 2 eV below $E_\mathrm{F}$, respectively.
It had been interpreted that the 4$f^0$ states came from the localized nature of 4$f$ electrons, and the peak intensity ratio 4$f^0$/4$f^1$ had been taken as a measure of the 4$f$-electron localization.
When the bulk sensitivity is enhanced by increasing the photon energy from VUV to SX, the ratio was suppressed significantly in SX-PES experiments.
In addition, the SX-PES experiments also have revealed the bulk electronic states for the transition-metal oxides \cite{Sekiyama_PRL, Mo_PRB}.
The SX-PES studies of the 4$f$ and 3$d$ electron systems have thus provided us with important information about the bulk electronic states and help us to understand the intrinsic nature of strongly correlated electron systems.
It naturally follows that there would be a similar situation in the 5$f$ electron systems, namely, that there is a possibility that the U 5$f$ bulk electronic states of UX$_\mathrm{C}$ have been hardly detected in the VUV-PES experiments.
However, there has no report in which the surface states of uranium compounds are treated in detail.
As another merit of SX-PES experiments, it is expected that the bulk U 5$f$ states can be experimentally identified because the photoionization cross-section of the U 5$f$ orbital relative to the other orbitals is large in the SX energy region than in the VUV energy region.

In this paper, we have measured SX-PES spectra of uranium monochalcogenide US, and revealed the intrinsic bulk U 5$f$ electronic states. 
In SX-PES, the {\it B} structure, which had been previously considered as the localized character of the U 5$f$ electrons, was never observed in the valence-band spectrum of US.
Instead, the SX-PES valence-band spectra showed good agreement with that of the band-structure calculations in the local-density approximation (LDA). 
The results suggest that US should be an itinerant-electron ferromagnet.
The present study is the first demonstration for uranium compounds that the surface states, which hinder us seriously from understanding the intrinsic electronic states, are clearly identified.

Single-crystalline samples of US were grown by the Bridgman method, starting from a uranium metal of 99.95 {\%} and chalcogenide of 99.999 {\%}.
The SX-PES measurements were performed at the beamline BL23SU of SPring-8. 
The temperature was set to 20 K and the energy resolution ($\Delta$E) was 80 - 100 meV at the excitation photon energies of $h{\nu}$ = 400 - 800 eV.
Clean surfaces of the samples were obtained by cleaving them at 20 K.
[111] surfaces, which are the easiest cleavage planes, were exposed.

\begin{figure}
\includegraphics[scale=0.5]{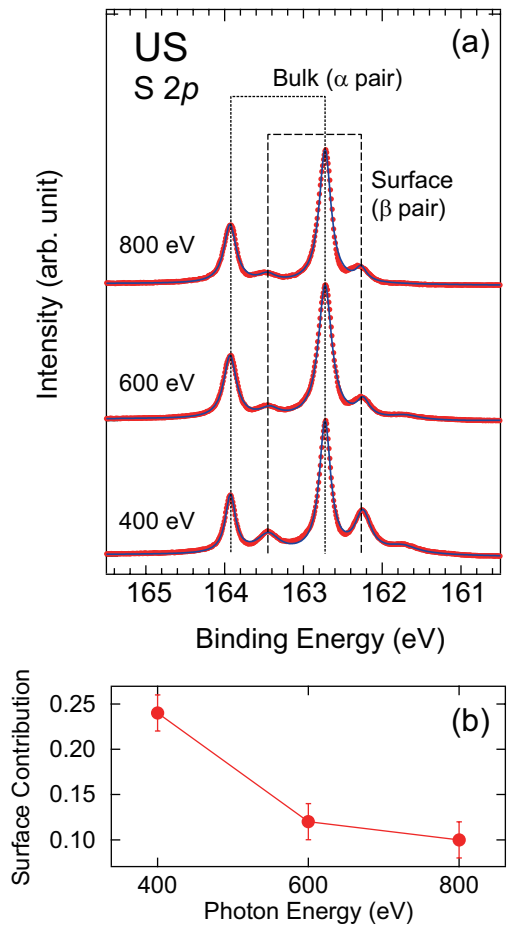}
\caption{\label{fig:Fig1} (Color online) S 2$p$ core-level PES experiment of US. (a) Photon-energy dependences of core-level PES spectra. Red points are experimental data and blue lines are fitted curves. (b) Photon-energy dependences of the surface contribution from the curve fitting as shown in Fig.1 (a). The procedure of the estimation of the surface contribution is described in text.}
\end{figure}

We performed core-level PES experiment to investigate whether surface states exist or not.
Figure 1 (a) shows the photon-energy dependences of the S 2$p$ core-level PES spectra.
Since the crystal structure of UX$_\mathrm{C}$ is the NaCl type, there is only one crystallographic site for the S atoms.
Therefore, the S 2$p$ core-level PES spectra must have only one pair of spin-orbit splitting peaks of the S 2$p_{1/2}, _{3/2}$.
All the S 2$p$ core-level spectra for US show two pairs ($\alpha$ at deeper $E_\mathrm{B}$ and $\beta$ at lower $E_\mathrm{B}$) with clear chemical shifts.
The S 2$p$ core-level spectra have the $\alpha$ pair (peaks at $E_\mathrm{B}$ = 164.0 and 162.8 eV) and the $\beta$ pair (peaks at $E_\mathrm{B}$ = 163.5 eV and 162.2 eV) as shown in Fig. 1 (a).
The intensity of the $\beta$ pair is suppressed as the photon energy increases from 400 to 800 eV.
This indicates that the $\beta$ pair intensity is related to the contribution from the surface states.
There is an additional peak at $E_\mathrm{B}$ = 161.8 eV (see the spectrum at $h{\nu}$ = 400 eV).
We confirmed that the intensity of the additional peak was strong when the sample surface was oxidized by checking the O 1$s$ core-level PES intensity.
This indicates that the $\beta$ pair does not come from the surface oxidation.
All the measurements were performed at a sample position where the additional peak was suppressed and the O 1$s$ core-level PES intensities disappeared.
Although the intensity of the additional peak was decreased as the photon energy increased, the photon-energy dependence of the additional peak is much smaller than that of the $\beta$ pair.
Therefore, the surface contribution can be estimated by the integrated intensity of the $\beta$ pair relative to the total spectral intensity by multi-peak fitting with Gaussian functions (blue solid lines) as shown in Fig. 1 (a).
The surface contribution is estimated to be 24, 12 and 10 {\%} at $h{\nu}$ = 400, 600 and 800 eV with an error $\pm$ 2 {\%}, respectively.
Fig. 1 (b) shows the photo-energy dependence of the surface contribution obtained by the above procedure.
The surface contribution is reduced significantly as the photon energy increases.
Therefore, we can deduce that the previous VUV-PES spectra of US are strongly affected by the surface states.
From the photon-energy dependence of the S 2$p$ core-level spectra, it can be expected that the surface contribution of the valence-band spectrum of US at $h{\nu}$ = 800 eV is suppressed by less than 10 {\%}, because the photoelectron energy for the valence bands is 160 eV higher than that for the S 2$p$ core levels.

\begin{figure}
\includegraphics[scale=0.55]{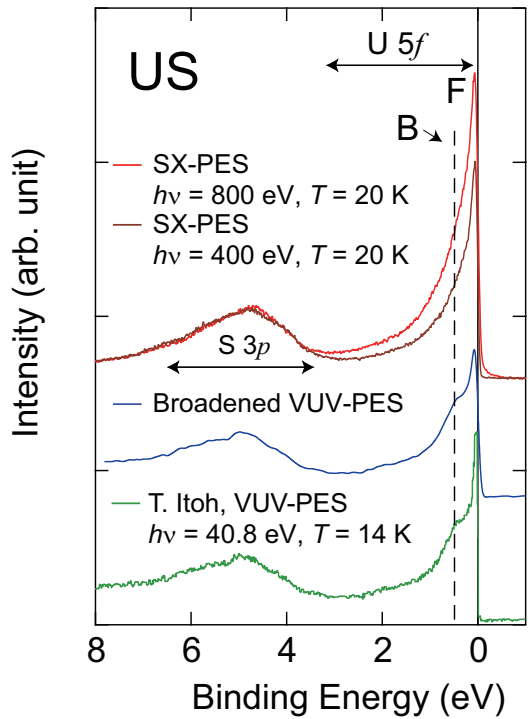}
\caption{\label{fig:Fig2} (Color online) Photon-energy dependences of the SX-PES valence-band spectra of US. Previous VUV-PES valence-band spectrum \cite{Ito_DoctorThesis} and the broadened VUV-PES spectrum by a Gaussian function ($\Delta$E = 100 meV) are also shown for comparison. These spectra are normalized by the intensity of the S 3$p$ states.}
\end{figure}

Next, we performed valence-band SX-PES experiment in order to extract bulk dominant electronic structures. 
Figure 2 shows the photon-energy dependence of the angle-integrated SX-PES spectra of US.
The valence-band spectra at $h{\nu}$ = 40.8 eV by Itoh \cite{Ito_DoctorThesis} are also shown for comparison.
The broadened VUV-PES spectrum are obtained by convoluted with a Gaussian function ($\Delta$E = 100 meV at $h{\nu}$ = 800 eV).
As the photon energy increases from 400 to 800 eV, the intensity near $E_\mathrm{F}$ is enhanced significantly relative to the intensity of the structures at $E_\mathrm{B}$ = 3 - 7 eV.
According to the photoionization cross-sections \cite{Yeh_1985}, the enhanced structures from $\sim$ 3.5 eV to $E_\mathrm{F}$ and the structures around 5 eV can be ascribed to the U 5$f$ and S 3$p$ states, respectively (see Fig. 3).
The U 5$f$ states exist not only in the vicinity of $E_\mathrm{F}$, but are also spread over the comparatively wide energy range of $\sim$ 3 eV.
We emphasize that only the {\it F} structure with a sharp peak near $E_\mathrm{F}$ is observed, while the {\it B} structure observed at $E_\mathrm{B}$ = 0.5 - 1.0 eV in the previous VUV-PES studies disappears \cite{Reihl_Common128_1987, Ito_DoctorThesis}. 
It should be noted that the {\it B} structure clearly survives in the broadened VUV-PES spectrum.
Therefore, the absence of the {\it B} structure in the SX-PES spectra is not due to the experimental resolution.
Since the {\it B} structure disappears/appears depending on the photon energies, the {\it B} structure is considered to be derived mainly from the electronic states at the surface.

\begin{figure}
\includegraphics[scale=0.55]{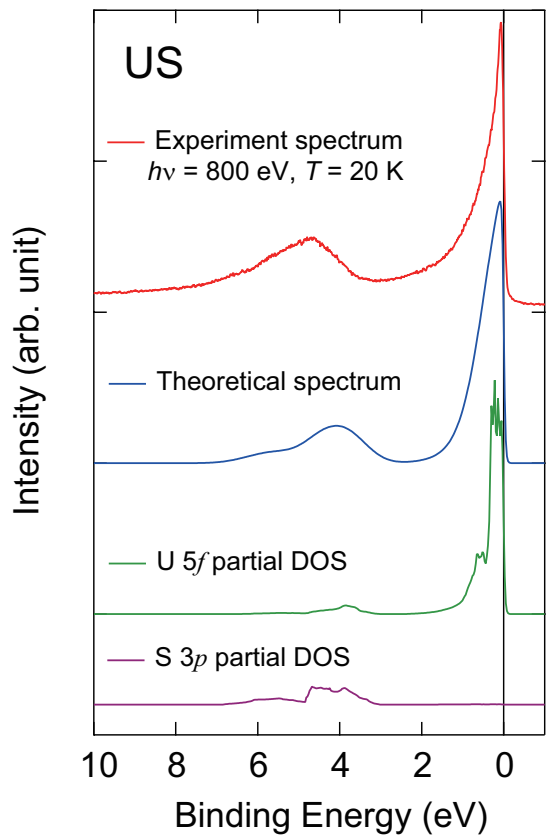}
\caption{\label{fig:Fig3} (Color online) Comparison of the experimental valence-band spectrum at $h{\nu}$ = 800 eV and the theoretical valence-band spectra of US deduced from the band-structure calculation taking into account the photoionization cross-sections at $h{\nu}$ = 800 eV. The S 3$p$ and U 5$f$ partial DOS are also displayed.}
\end{figure}

Figure 3 shows quantitative comparison between the experimental valence-band spectrum at $h{\nu}$ = 800 eV and the valence-band density of states (DOS) in the ferromagnetic states calculated by the relativistic linearized-augmented-plane wave method within the LDA \cite{Yamagami_JPSJ_1998}.
The theoretical DOS at $h{\nu}$ = 800 eV has been synthesized by taking into account the photoionization cross-sections each of the orbital \cite{Yeh_1985} and then by convoluting with a Gaussian function ($\Delta$E = 100 meV).
The valence band mainly consists of the S 3$p$ and U 5$f$ states as shown in Fig. 3.
Although the S 3$p$ and U 5$f$ states are weakly hybridized with each other, these states are basically separated in energy.
The S 3$p$ partial DOS near $E_\mathrm{F}$ is very small and the spectral weight near $E_\mathrm{F}$ comes from the U 5$f$ states.
The entire valence-band structures from the band calculation are very similar to the experimental spectrum, though the experimental S 3$p$-band positions are shifted a little deeper $E_\mathrm{B}$ than the theoretical ones.
Thus, the experimental spectrum of US can be reproduced by the band-structure calculation.
The band width of the U 5$f$ states of US of the experimental spectrum is somewhat narrower than that of the theoretical spectrum.
This indicates that an electron-correlation effect to the U 5$f$ electrons exists.
Therefore, we can conclude that the electronic states of US are described basically by the itinerant-band picture of the U 5$f$ electrons, indicating that US is an itinerant-electron ferromagnet.
This result is consistent with the indication that other uranium monochalcogenide UTe is an itinerant-electron ferromagnet from the temperature-dependent VUV-PES study \cite{Durakiewicz_PRL_2004}.
From the present study, we can conclude that the {\it B} structure of US in the previous VUV-PES studies, which has been considered as evidence for the localized nature of the U 5$f$ electrons and regarded as a fingerprint of the ``dual model", is originated from the surface contribution to the valence-band spectra.
In future we have to investigate whether the systematic change of the {\it B} structure from US to UTe comes from the intrinsic bulk state or not.
The present finding gives a new and clear guideline for understanding the 5$f$ states not only in uranium monochalcogenide, but also in other uranium compounds and/or actinide compounds.

In conclusion, we have investigated the electronic structure of US using SX-PES.
We could extract the intrinsic bulk electronic structure from the photon-energy dependence of the valence-band and core-level spectra.
In the valence-band spectra by SX-PES, there is no {\it B} structure that has been attributed to the localized U 5$f$ electrons in the previous VUV-PES studies.
The valence-band spectra can be explained basically by the band-structure calculation, thus indicating that the itinerant picture of U 5$f$ electrons is more appropriate to understand the electronic structures of US and hence the magnetic properties of uranium monochalcogenides.
The present study is the first report which points out the existence of the surface electronic states of uranium compounds using SX-PES experiments.
The finding will give important insight and great impact on the field of strongly correlated electron systems as well as of actinide material science.

This work was performed under the Proposal No. 2007B3822 at BL23SU of SPring-8 and supported by the Grant-in-Aid for Scientific Research on Innovative Areas "Emergence of Heavy Electrons and Their Ordering" (No.20102003) from the Ministry of Education, Culture, Sports, Science, and Technology, Japan.

\end{document}